\documentclass[useAMS,usenatbib]{mn2e}
\usepackage{color}
\usepackage[colorlinks]{hyperref}
\usepackage{amssymb}
\usepackage{amsmath}
\usepackage{graphicx}
\usepackage{xspace}
\usepackage{url}
\def\fermi{\textit{Fermi}\xspace}
\def\flat{\textit{Fermi}/LAT\xspace}
\def\xrt{\textit{Swift}/XRT\xspace}
\def\xmm{\textit{XMM-Newton}\xspace}
\def\cha{\textit{Chandra}\xspace}

\def\sn{SN~1987A\xspace}
\def\dorc{30~Dor~C\xspace}
\def\rxj{RX~J0536.9-6913\xspace}
\def\hn{Honeycomb nebula\xspace}

\begin{document}
	\title[Evidence for GeV brightening of SN 1987A]{Evidence for recent GeV brightening of the SN~1987A region} 
	\author[D. Malyshev et al]{D. Malyshev$^{1}$, G. P\"uhlhofer$^{1}$, A.Santangelo$^{1}$, J. Vink$^{2,3,4}$  \\ 
		$^{1}$ Institut f{\"u}r Astronomie und Astrophysik T{\"u}bingen, Universit{\"a}t T{\"u}bingen, Sand 1, D-72076 T{\"u}bingen, Germany\\
		$^2$ GRAPPA, University of Amsterdam, Science Park 904, 1098 XH Amsterdam, Netherlands\\
		$^3$ API, University of Amsterdam, Science Park 904, 1098 XH Amsterdam, Netherlands \\
		$^4$ SRON Netherlands Institute for Space Research, Utrecht, The Netherlands
	}


\maketitle
\label{firstpage}
\begin{abstract}
	{
    We report on a recent (2016--2018) enhancement of the GeV emission from the \sn region as observed with \flat. The observed signal is characterised by a power-law spectrum with a slope of $2.1\pm 0.2$  and is detected only at energies $\gtrsim 1$~GeV. The \flat data constrain the position of the signal to within $0.15^\circ$ around \sn. Although a recent increase in the gamma-ray emission from \sn seems to be a natural explanation for the detected emission, given the youth of the source and its rapid evolution, the \flat location also overlaps with several other potential gamma-ray sources: \dorc, the \hn, \rxj, and a hypothetical, previously unknown transient source. We argue that multiwavelength observations of the region performed during the next few years can clarify the nature of the signal and encourage such observations. We also present upper limits on the time-averaged flux of \sn based on $10$~years of \flat exposure, which can be used to better constrain the particle acceleration models of this source.\newline
	}
\end{abstract}
\section{Introduction}
\label{sec:intro}
\sn, which occurred on Feb. 23rd 1987, is a naked-eye core collapse supernova located in the Large Magellanic Cloud (LMC). Its brightness and relatively small distance ($\sim 50$~kpc) allowed to study the progenitor star and the explosion in great details. The consequent evolution of the supernova remnant has been monitored with gradually improving instruments over the last 30~years. Nowadays, \sn is a laboratory for fundamental studies covering a broad range from diffusive shock particle acceleration~\citep{ball92,berezhko11,berezhko15} and modeling of the expansion of supernova remnants~\citep{potter14} to sterile neutrinos~\citep{arguelles16} and axions~\citep{payez15}.

The spatial and spectral evolution of \sn has been traced from the radio to the hard X-ray band, see e.g.~\citet{zanardo17,chandra_sn1987a_2017,boggs15,reynolds15}, and \citet{mccray16} for a recent review.
The environment of \sn is very complex, the evolution takes place in an hour-glass shaped cavity bound by a dense ring in the equatorial plane.  Most of the X-ray emission is thermal, originating
from the shocked material in the dense ring, through which the blast wave was propagating until recently  \citep{chandra_sn1987a_2016}.  

The hydrodynamic model proposed by \citet{berezhko11,berezhko15} suggests efficient cosmic ray proton acceleration at the shock of \sn. This leads to a strong shock modification and consequently to a steep spectrum of accelerated electrons, which allows to explain the steep radio spectrum of the source.
Interacting with the surrounding medium, mainly the material in the ring, the accelerated protons are expected to produce a significant level of GeV-TeV $\gamma$-ray emission in $\pi^0$-decay processes. The changes of the medium's density by a factor of a few should lead to corresponding changes of the GeV-TeV flux of the source. 

The model by \citet{berezhko11,berezhko15} predicts a gradual brightening of \sn in the GeV-TeV band by a factor of 2 in the 2010 -- 2030 epoch with a peak emission that corresponds to the time close or right after the shock has completely penetrated the ring.
Contrary to this, no VHE emission has been detected from \sn so far despite several dedicated analyses of the region in the GeV~\citep{lat_lmc1,lat_lmc} and TeV~\citep{30dorc_hess} bands.

In the following, we present the results of an analysis of more than 10~years of monitoring data of \sn obtained with \flat in the $0.1$~GeV -- 3~TeV energy band, which can be used to constrain evolutionary models of the object. 
The analysis resulted in a detection of GeV emission from the \sn region, for the last two years of the observations (2016-2018). We accompany our analysis with X-ray \xmm and \xrt data and discuss the possible multiwavelength counterparts of GeV signal. 

\section{Data analysis}
\subsection{\flat data analysis}
\label{sec:lat_data_analysis}
\flat data that were selected for the analysis which is presented in this paper cover more than 10 years (Aug. 2008 to Dec. 2018). For the main analysis, we used the latest available \texttt{fermitools} with P8\_R3 response functions (\texttt{CLEAN} photon class)\footnote{See the \href{https://fermi.gsfc.nasa.gov/ssc/data/analysis/documentation/Cicerone/Cicerone_LAT_IRFs/c }{description of the \flat response functions} }. 

To extract the spectra for all cases presented below, we performed a standard binned likelihood analysis of the region around \sn. The spectral analysis is based on the fitting of the spatial / spectral model of the sky region around the source of interest to the data. The region-of-interest considered in the analysis is a circle of 14 degrees radius around \sn. The model of the region includes all sources from the FL8Y catalogue\footnote{See the \href{https://fermi.gsfc.nasa.gov/ssc/data/access/lat/fl8y/FL8Y_description_v8.pdf}{source list description document}} as well as components for isotropic and galactic diffuse emission given by the standard spatial/spectral templates \texttt{iso\_P8R3\_CLEAN\_V2.txt} and \texttt{gll\_iem\_v06.fits}. 

The spectral template for each FL8Y source in the region was selected according to the catalogue model. The normalisations of the sources were considered to be free parameters during the fitting procedure.  In order to avoid possible systematic effects, we have also included the FL8Y sources located up to $10^\circ$ beyond the ROI into the model, with all parameters fixed to their catalogue values. Following the recommendation of the \flat collaboration, we performed our analysis enabling energy dispersion handling. All upper limits presented in this work were extracted for points with test-statistics (TS; \citealt{mattox96}) $TS<4$  with the \texttt{UpperLimits} python module provided with the \fermi/LAT software and correspond to a 95 per cent ($\simeq 2\sigma$) confidence level.

\begin{figure}
\includegraphics[width=1.\linewidth]{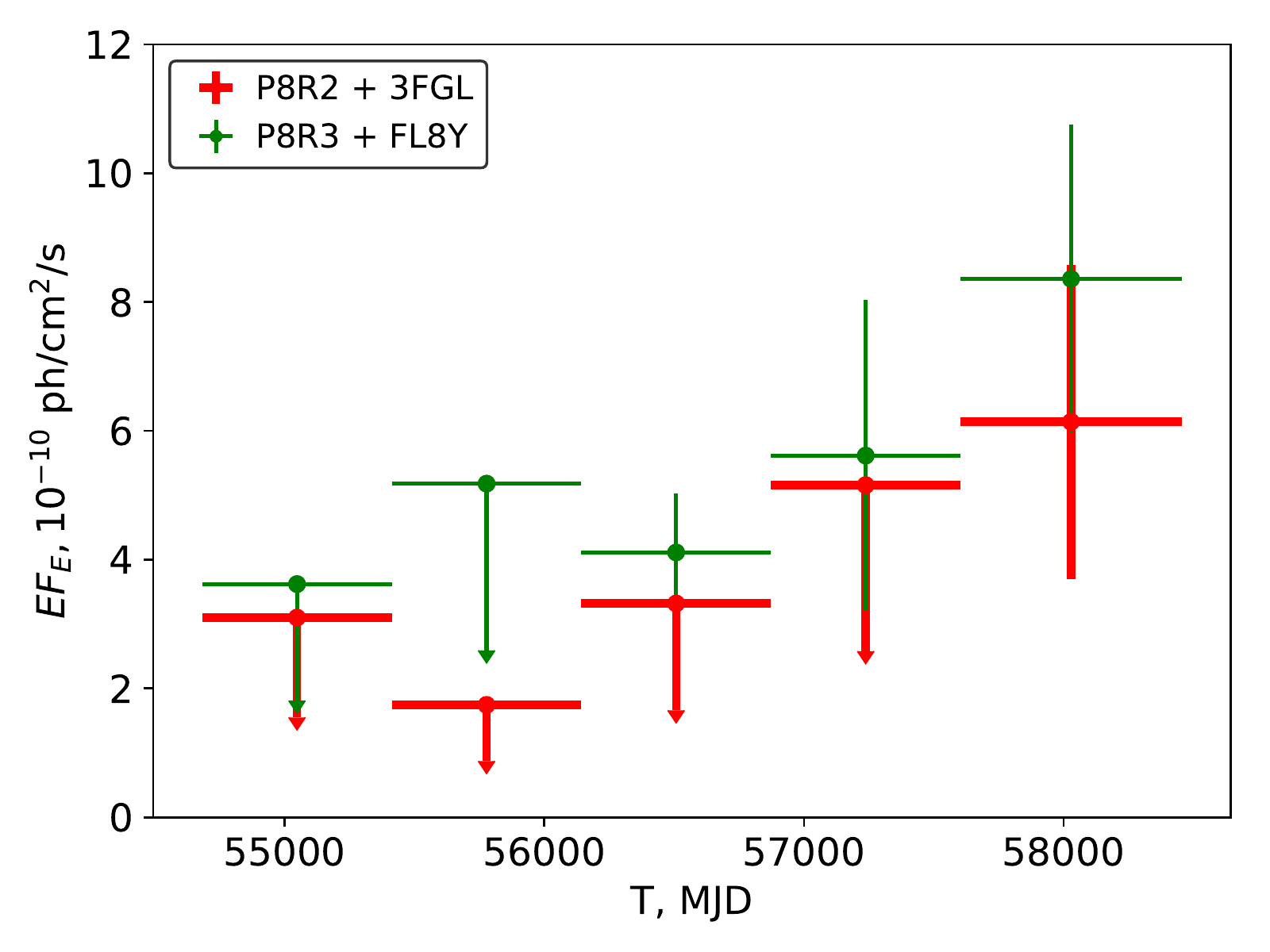}
\caption{ \flat lightcurve of the \sn region at energies 1-100~GeV. Each point corresponds to 2~years of data. Red points illustrate the results obtained with the most recent analysis chain (P8R3 IRFs and sources from FL8Y catalogue). Green points stand for the cross-check analysis chain with P8R2 IRFs (using the 3FGL source catalogue), see text for details.}
\label{fig:lc}
\end{figure}

At the initial stage of the analysis, we performed the spectral analysis as described above for five 2-year-long time bins at energies above 1~GeV, with the goal to build a light-curve for \sn. We found that the source was detected with $TS>9$ (i.e.\ above $3\sigma$ significance)  in the last two time bins (Aug. 2014 -- Aug. 2016 and Aug. 2016 -- Dec. 2018). The light-curve is shown with green symbols in Fig.~\ref{fig:lc}, the test-statistics of the last two bins is $TS\approx 11$ and $TS\approx 25$ corresponding to $\sim 3\sigma$ and $\sim 5\sigma$ detection significance, respectively. During Aug. 2012 -- Aug. 2014 (the third point in the light-curve) the source is only marginally detected with $TS\approx 5$. Given the five considered time bins, a trial factor of 5 can be applied, which still yields a detection significance in the last time bin of $\sim 4.7\sigma$.

To cross-check the results, we also analysed the data calibrated with the P8\_R2 instrumental functions (as recommended for analysis by the \flat collaboration before the P8\_R3 IRFs were released in Dec.~2018) and used the 3FGL catalogue of \fermi sources for the model of the region. In this case, we also explicitly added  to the model of the region sources that were found by the \flat collaboration during the dedicated analysis of the LMC~\citep{lat_lmc}. 

The results of the cross-check analysis are shown in Fig.~\ref{fig:lc} with red symbols. These results are compatible with those of the main analysis, although with the P8\_R3 IRFs the source is detected already at earlier times and with slightly higher significance. This can be attributed to the improved background rejection of the P8\_R3 calibration. To be conservative, we focus only on the 2016-2018 detection of \sn which is confirmed with both analysis chains ($TS=19$ with the cross-check analysis).

\begin{figure*}
\includegraphics[width=1.\linewidth]{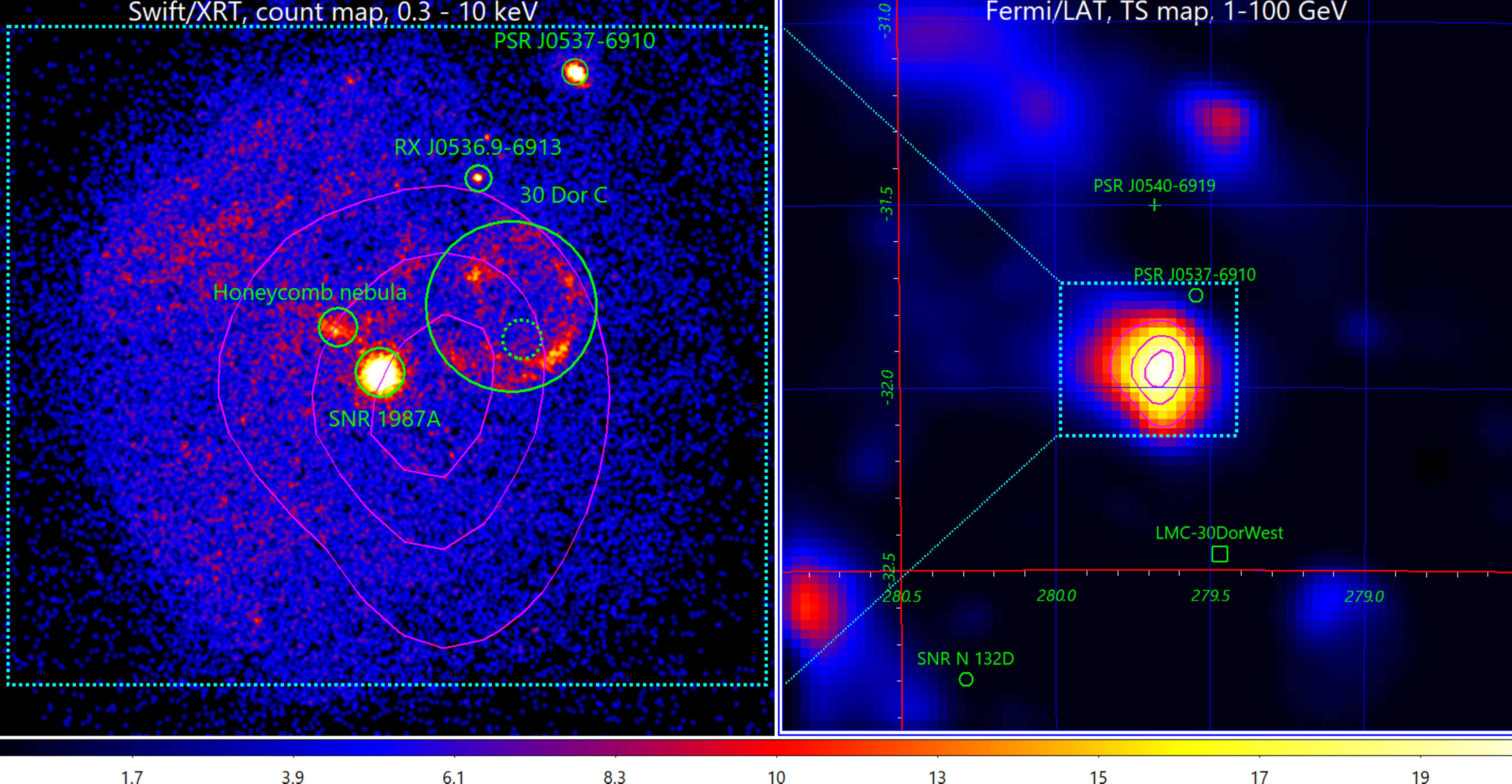}
\caption{\xrt 0.3--10~keV count map (left panel) and \flat test-statistics (TS) map (right panel) in the 1--100~GeV energy range, in galactic coordinates. Magenta contours on both maps are identical and indicate $1\sigma$, $2\sigma$, and $3\sigma$ confidence regions for the \flat location of the emission. Cyan square in the right panel indicates the scale of the left panel. The positions of FL8Y \flat catalogue sources subtracted from the TS map during the analysis are shown in the right panel with green symbols. The dashed circle in the left panel illustrates HESS localization of TeV signal from \dorc region~\citep{30dorc_hess} }
\label{fig:ts_map}
\end{figure*}

In the following, we refer to the period of the detection of the GeV signal (Aug. 2016 -- Dec. 2018) as ``high-flux period'', and as ``low-flux period'' to the whole time period before the signal detection (Aug. 2008 -- Aug. 2016). The test-statistics map for the high-flux period, built in the 1-100~GeV energy band, is presented in the right panel of Fig.~\ref{fig:ts_map}. All sources included into the model, except \sn, were taken into account and thus do not appear in this map. The magenta contours indicate the $1\sigma, 2\sigma, 3\sigma$ confidence regions for the position of the non-modeled source. 

\begin{figure*}
\includegraphics[width=0.47\linewidth]{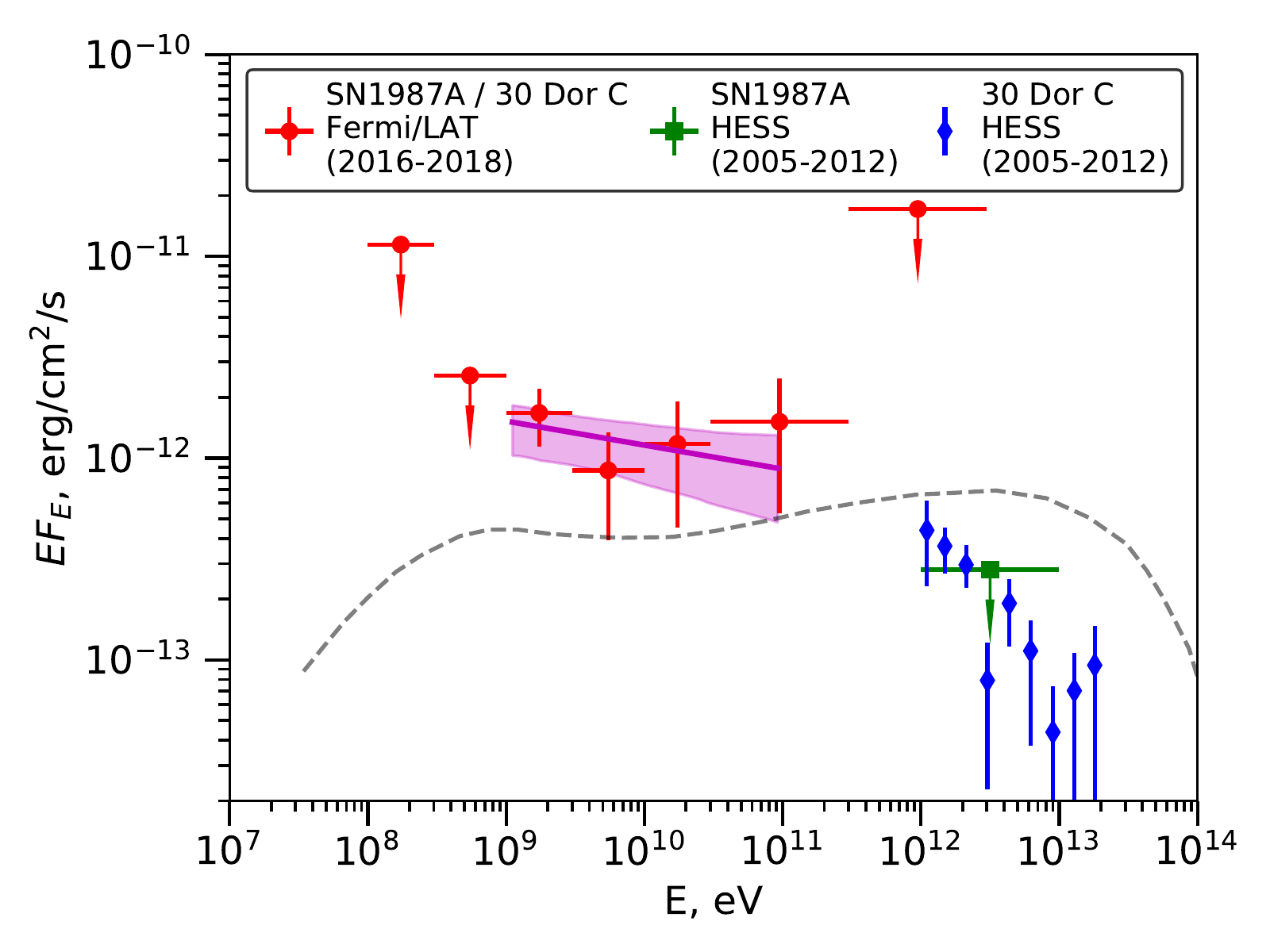}
\includegraphics[width=0.47\linewidth]{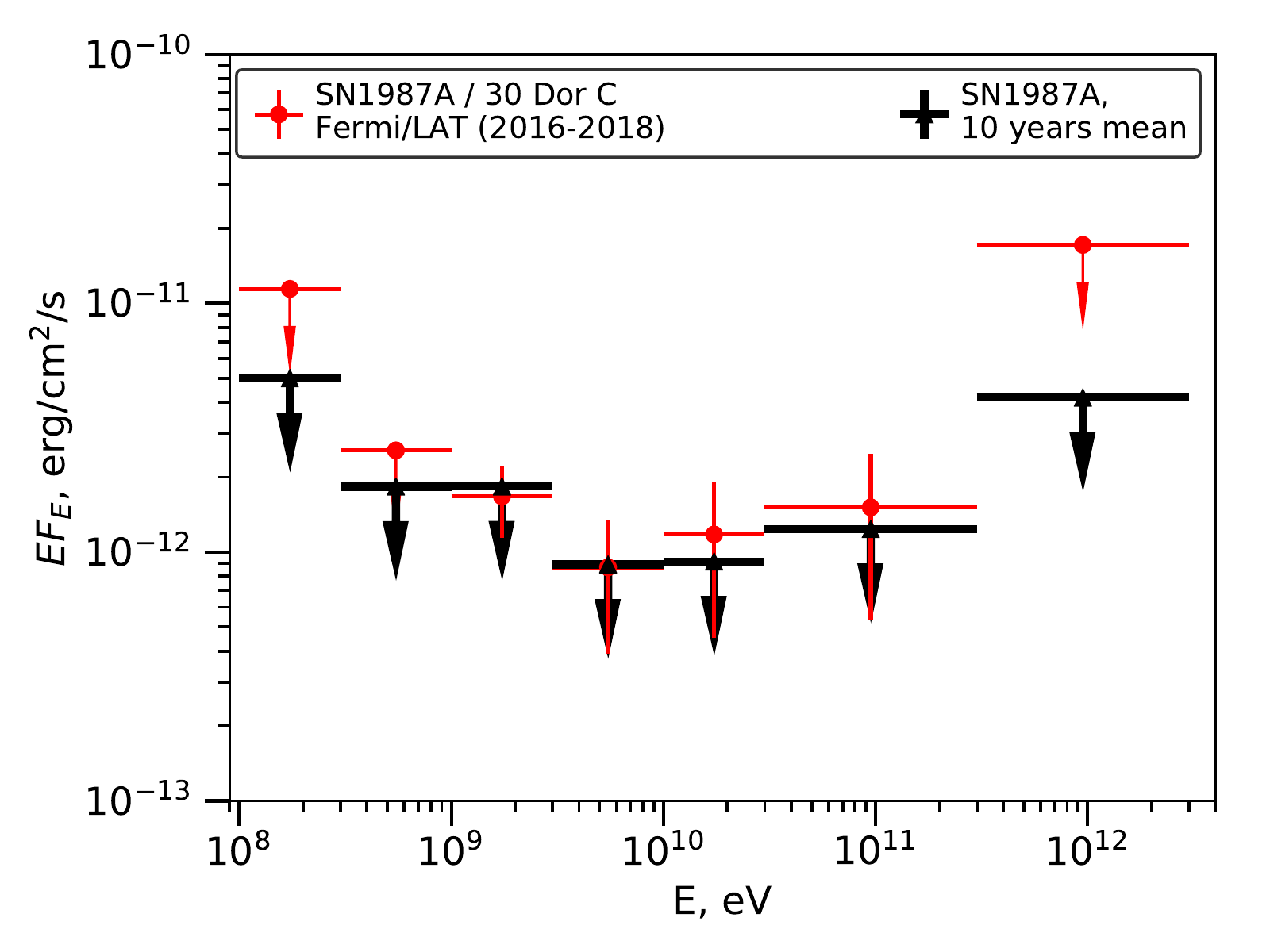}
\caption{\textit{Left:} \flat spectrum of the \sn region during 2016--2018. The magenta line and the shaded region illustrate the best-fit of the data in the 1-10~GeV energy range with a simple power-law model the and corresponding $1\sigma$ confidence range for the fit, respectively. The black dashed line illustrates the prediction of the~\citet{berezhko11} model for the flux of \sn in the 2030 epoch. The green upper limit and the blue points show the HESS results for \sn and \dorc obtained in the 2012-2015 period~\citep[adapted from][]{30dorc_hess}.\newline\textit{Right:} Flux upper limits of \sn, from 10~years of \flat data. Red points illustrate the spectrum of \sn during 2016--2018.}
\label{fig:spectrum}
\end{figure*}

Besides \sn, the $3\sigma$ contour includes a number of possible counterpart sources, including \dorc, the \hn, and \rxj. The left panel of Fig.~\ref{fig:ts_map} presents the \xrt map of the region with marked positions of possible counterpart sources. 
The spectrum of the signal observed during the high-flux period is characterised by a power-law spectrum with a slope of $\Gamma=2.1\pm 0.2$ at energies 1-100~GeV, see Fig.~\ref{fig:spectrum}. 

In our analysis we also considered the time-averaged \flat data and explicitly calculated flux upper limits for each energy interval. The results for the spectral upper limits are shown in Fig.~\ref{fig:spectrum}, right panel.

\subsection{X-ray data}
We accompanied the analysis with historic \xmm and \xrt observations including the data from recent (May 2018) dedicated \xrt TOO observation of the region. \xrt data was reprocessed and analysed with the \texttt{xrtpipeline v.0.13.4} and \texttt{heasoft v.6.22} software packages, following recommendations of the \xrt team\footnote{See e.g. the \href{https://swift.gsfc.nasa.gov/analysis/xrt_swguide_v1_2.pdf}{\xrt User's Guide}}. To reduce potential \xrt pileup problems for a bright soft source, we extracted the spectrum of \sn from an annulus with an excluded central region of $8''$ radius. 

The considered \xmm data consists of three observations (ObsIds: 0763620101, 0783250201, 0804980201 ) performed just before and during the high GeV flux period (Nov. 2015, Nov. 2016 and Oct. 2017 correspondingly). EPIC-pn spectra were extracted using SAS version 17.0.0. 
All spectra were extracted from a $30''$-radius circular region centred on the source. The background spectra were extracted from a nearby source-free region. The spectra of \sn were additionally corrected for pileup, excluding the central $7.5''$-radius region, using the method described in the relevant SAS data analysis thread\footnote{See \href{https://www.cosmos.esa.int/web/xmm-newton/sas-thread-epatplot}{``How to evaluate and test pile-up in an EPIC source''} thread} and modelled with a thermal, three-component plane-parallel shock model (\texttt{vpshock} in XSPEC) similar to~\citet{maggi12}.

The spectral analysis was performed with \texttt{XSPEC v.12.10.0c}. The evolution of the 0.3--10~keV flux of \sn and nearby X-ray sources within \flat localization uncertainties is shown in Fig.~\ref{fig:all_swift}.

\section{Results and discussion}
\label{sec:discussion}
The \flat localisation area of the signal with a radius of $\sim 0.15^\circ$ (see Fig.~\ref{fig:ts_map}) contains a few possible multiwavelength counterparts, including \sn, \dorc, the \hn, and \rxj which is just barely beyond the $3\sigma$ localisation contour. The poor statistics of the GeV data also does not allow to exclude the presence of a transient source active for some period during Aug.~2016-2018. 

Assuming that the source of the signal is located in the LMC, we estimate the luminosity of the emission during the high-flux period to be
\begin{equation}
L_{\gamma} \sim 5\cdot 10^{35} \left( \frac{D}{50\mbox{kpc}} \right)^2\mbox{erg/s}
\end{equation}
This luminosity is characteristic for supernova remnants detected in the GeV band~\citep[see e.g.][]{dermer13}, which marginally supports a supernova remnant (\sn or \hn) origin of the GeV emission.
In the following, we discuss the details of the possible counterparts to the GeV signal. 

\subsection{\sn}
\sn is the most likely and intriguing potential counterpart for the $\gamma$-ray emission, since a GeV-TeV brightening of the source is expected around 2010--2030~\citep{berezhko11,berezhko15}. The GeV flux observed in 2016-2018 is higher than the one predicted by the model, see the dashed curve in Fig.~\ref{fig:spectrum}. However, \sn is developing in a rather complex environment with a variety of densities and with varying shock speeds. In such conditions it can be difficult to accurately predict the cosmic ray acceleration efficiency and the gamma-ray emissivities from the first principles. A rather uncertain renormalisation factor of 5 applied by~\citet{berezhko11} to their modeled $\gamma$-ray spectrum, indicates the level of uncertainty of the prediction and shows that the measured spectrum is actually reasonably consistent with the model.

\begin{figure}
\includegraphics[width=1.\linewidth]{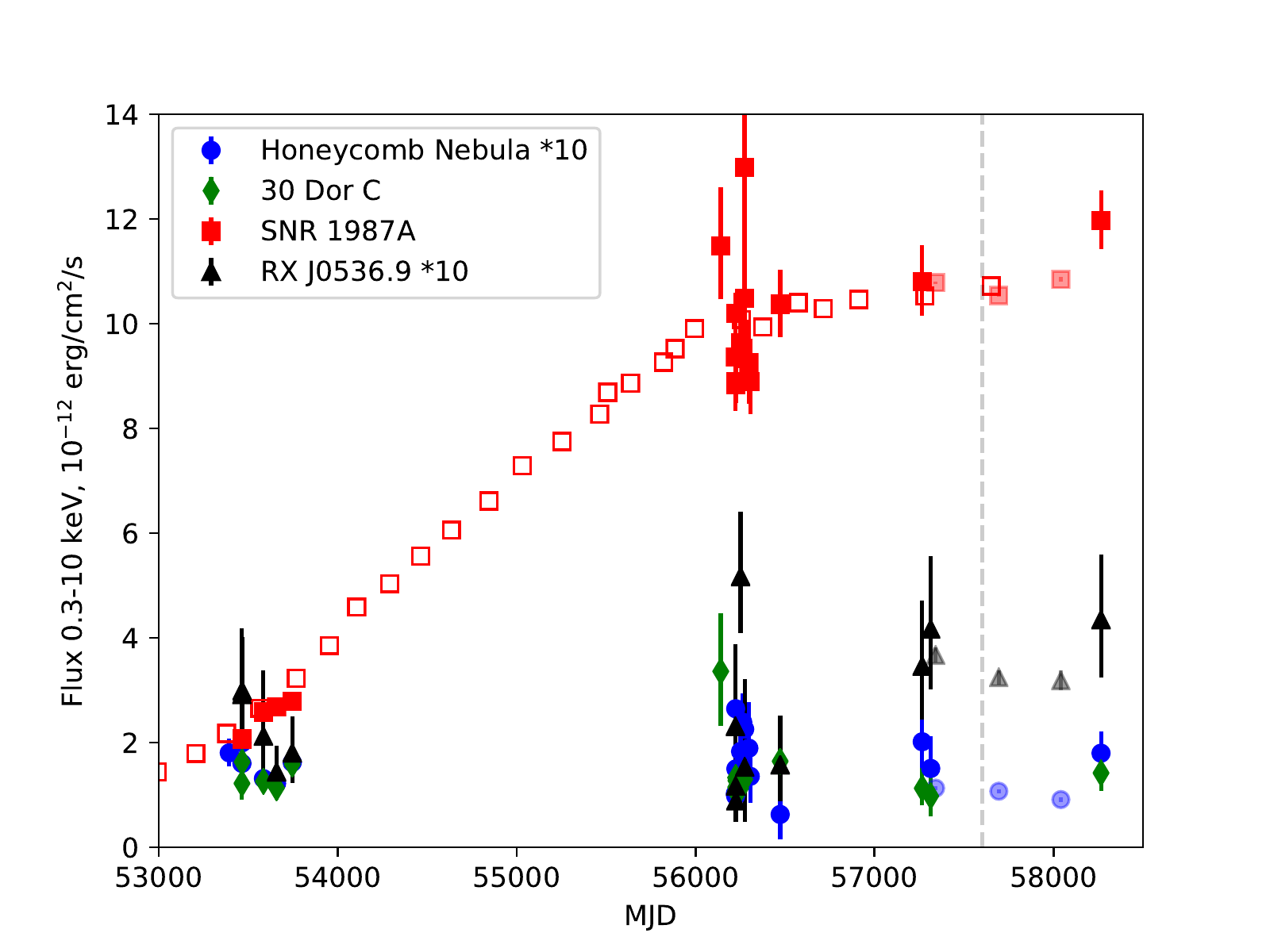}
\caption{\xrt(solid symbols) and \xmm (half-transparent symbols) fluxes of all sources within the $3\sigma$ \flat positional confidence region of the GeV excess. The empty boxes illustrate Chandra measurements of the \sn flux reported by~\citet{chandra_sn1987a_2017}, rescaled to 0.3-10~keV. The vertical dashed line illustrates the start of the high GeV-flux period (MJD$>$57602).}
\label{fig:all_swift}
\end{figure}

The X-ray light-curve of \sn demonstrates a comparatively stable behavior, see Fig.~\ref{fig:all_swift} for the combined \xrt, \xmm and \cha light-curves (adapted from ~\citet{chandra_sn1987a_2016,chandra_sn1987a_2017}). An indication of a $\sim 30$\% flux increase can be seen in the \xrt data close to high GeV flux period. 

The absence of such variations in the sparser but higher quality \cha and \xmm data may point towards an instrumental (e.g.\ pileup) origin of this flux variation.
A stable X-ray flux of \sn during the high GeV-flux period can in general be understood if the gamma-ray emission from \sn is dominated by the $\pi^0$-decay component, as suggested also by~\citet{berezhko11,berezhko15}. 

On the other hand, if confirmed, the increase of the X-ray flux would suggest a leptonic scenario of the emission production in \sn, but only if 
the increase of the X-ray flux  is not due to thermal emission, but caused by X-ray synchrotron emission of relativistic electrons accelerated at the \sn shock. This additional spectral component would come on top of the currently observed bright thermal emission, which would explain the relatively small changes of the X-ray flux. We would like to note that the statistics of the discussed \xrt observation does not allow to make any conclusions on the presence or absence of such a component.

Radio observations of \sn could serve as another potentially interesting way to probe the origin of the emission. For a leptonic scenario one would naturally expect a much higher increase of the radio flux in comparison to hadronic models. However, published radio data only cover the period before 2012~\citep{zanardo17} and cannot serve for this purpose. 

Another option is to exploit the slope of the \sn radio spectrum. The index $\alpha=0.74$~\citep{callingham16} implies rather different GeV-band slopes for the leptonic and hadronic scenarios, respectively. Assuming a synchrotron self-absorption model for the radio emission, the slope of the electron population is estimated to be $\Gamma_e = 2\alpha+1 = 2.48$. If the GeV emission has a leptonic (IC) origin, the GeV slope would therefore be  $\Gamma_l = (\Gamma_e+1)/2 = 1.74$. 
Within hadronic models, the GeV slope is $\Gamma_h=\Gamma_e = 2.48$ under the assumption that relativistic electrons and protons share the same index. The quality of the \flat data however does not allow us to firmly discriminate between the two slopes. Checking the difference between the best-fit log-likelihoods of the models with the spectral slope of \sn frozen to the corresponding values we explicitly verified that both indices are just marginally beyond the $1\sigma$ slope confidence range ($1.4\sigma$ for $\Gamma_h$ and $1.7\sigma$ for $\Gamma_l$).

In case the observed GeV emission is not connected to \sn, the reported flux from the region has to be interpreted as an upper limit on the flux from \sn, see Fig.~\ref{fig:spectrum} (right panel) for the upper limits based on $\sim 10$~years of \flat data. We would like to note that these limits by themselves are already constraining for the evolutionary models of this object.

\subsection{Other possible counterparts}
\paragraph*{\dorc} was classified as a superbubble by~\citet{mathewson85}. Superbubbles, created and powered by the collective action of stellar winds and SNRs, have for a while been considered to be possible cosmic ray accelerators~\citep{bykov01_acc,ferrand10}. After dedicated searches the Cygnus Cocoon~\citep{fermi_cyg11,bartoli14,neronov15,veritas_cyg} and \dorc \citep{30dorc_hess} were discovered to emit GeV-to-TeV $\gamma$-rays. The $\gamma$-ray emission from the shocks in superbubbles, however, is believed to be variable at $\gtrsim 10^5$~years timescales~\citep[see e.g.][]{bykov01_acc}, which challenges the association of the variable GeV signal with \dorc. 

The emission from individual young supernova remnants within the \dorc superbubble \citep[see e.g.][]{lazendic08, kavanagh15} may however be variable on much shorter timescales and could thus in principle be responsible for the observed GeV signal.

We would like also to note that the observed GeV emission could be also associated with certain small-scale structures in \dorc, e.g. bright X-ray knots, see~\citet{kavanagh19}. The origin of these knots is unclear but thought to be associated with enhanced emission in shock-cloud interaction regions~\citep{kavanagh19}. In this case the knot emission can be variable on time-scales of few years \citep{inoue09} and could potentially be associated with the observed GeV emission.

\paragraph*{The \hn} is a relatively faint X-ray source located spatially close to \sn, see Fig.~\ref{fig:ts_map}. Discovered in the early 90's~\citep{wang92}, the \hn is argued to be a peculiar supernova remnant developing in a shell of material of an older explosion~\citep{meaburn10}. The supernova remnant origin of this source and the peculiar conditions in which its shell is expanding may suggest that the GeV emission of the \hn is similar to that of \sn. This emission can be variable on a $\sim 20$~yrs time scale and could be detected by \flat. There is also the possibility of a microquasar nature of the \hn as proposed by~\citet{meaburn10}. Microquasars are known to be variable GeV emitters with spectra continuing to at least tens of GeVs~\citep[see e.g.][]{cygx3_fermi,cygx3,cygx1_1,cygx1_2}. These objects are detected in the GeV band only in a certain state, which could explain the variability of the GeV signal detected from the \sn region. The absence of clear X-ray variability of the \hn, however, supports the supernova remnant nature of this source. We nevertheless formally cannot exclude a switching of the \hn to another X-ray state for a significant duration during the period of the measured high GeV flux.

\paragraph*{\rxj/Transient source.}
\rxj is an X-ray source discovered by ROSAT~\citep{manami00}. It was argued to be a background AGN by~\citet{haberl01}, based on the high hydrogen column density derived from \xmm observations and a characteristic radio spectral index.

AGNs are known to be strongly variable sources with spectra extending to GeV and TeV energies. Although the \flat signal is centered $\sim 3\sigma$ away of this source we cannot strongly exclude that the observed GeV emission originates from a flaring state of this source. The recent X-ray \xrt observations do not exhibit an increase of the flux of \rxj and do not reveal any other high-flux transient source in the region. This allows us to constrain the variability timescale to be months to years, which is consistent with AGN variability timescales. 

\section{Conclusion}
\label{sec:conclusion}

The statistics of the recent observations of \flat of the \sn region does not permit to draw a firm conclusion on the counterpart of the detected GeV emission from the region and cannot exclude an extragalactic transient source origin of the signal. Still, the data show an enhancement of GeV emission at the $\sim 5\sigma$ statistical significance level for the last two years of observations, for which \sn provides the most natural origin of the emission, given its youth and rapid overall evolution. We argue that multiwavelength monitoring of the region within the next few years will allow a conclusion whether the observed GeV excess originates from a variable transient source or from a quasi-steady source like \sn. We encourage detailed studies of the region to identify the counterpart to the enhanced \flat emission.
\newline

\textit{Acknowledgements}. The authors acknowledge support by the state of Baden-W\"urttemberg through bwHPC. This work was supported by the Carl-Zeiss Stiftung through the grant ``Hochsensitive Nachweistechnik zur Erforschung des unsichtbaren Universums'' to the Kepler Center f{\"u}r Astro- und Teilchenphysik at the University of T{\"u}bingen. This work made use of data supplied by the UK Swift Science Data Centre at the University of Leicester. We acknowledge the use of public data from the Swift data archive and thanks the entire Swift team for accepting and planning Target-of-Opportunity request.

\def\aj{AJ}%
\def\actaa{Acta Astron.}%
\def\araa{ARA\&A}%
\def\apj{ApJ}%
\def\apjl{ApJ}%
\def\apjs{ApJS}%
\def\ao{Appl.~Opt.}%
\def\apss{Ap\&SS}%
\def\aap{A\&A}%
\def\aapr{A\&A~Rev.}%
\def\aaps{A\&AS}%
\def\azh{AZh}%
\def\baas{BAAS}%
\def\bac{Bull. astr. Inst. Czechosl.}%
\def\caa{Chinese Astron. Astrophys.}%
\def\cjaa{Chinese J. Astron. Astrophys.}%
\def\icarus{Icarus}%
\def\jcap{J. Cosmology Astropart. Phys.}%
\def\jrasc{JRASC}%
\def\mnras{MNRAS}%
\def\memras{MmRAS}%
\def\na{New A}%
\def\nar{New A Rev.}%
\def\pasa{PASA}%
\def\pra{Phys.~Rev.~A}%
\def\prb{Phys.~Rev.~B}%
\def\prc{Phys.~Rev.~C}%
\def\prd{Phys.~Rev.~D}%
\def\pre{Phys.~Rev.~E}%
\def\prl{Phys.~Rev.~Lett.}%
\def\pasp{PASP}%
\def\pasj{PASJ}%
\def\qjras{QJRAS}%
\def\rmxaa{Rev. Mexicana Astron. Astrofis.}%
\def\skytel{S\&T}%
\def\solphys{Sol.~Phys.}%
\def\sovast{Soviet~Ast.}%
\def\ssr{Space~Sci.~Rev.}%
\def\zap{ZAp}%
\def\nat{Nature}%
\def\iaucirc{IAU~Circ.}%
\def\aplett{Astrophys.~Lett.}%
\def\apspr{Astrophys.~Space~Phys.~Res.}%
\def\bain{Bull.~Astron.~Inst.~Netherlands}%
\def\fcp{Fund.~Cosmic~Phys.}%
\def\gca{Geochim.~Cosmochim.~Acta}%
\def\grl{Geophys.~Res.~Lett.}%
\def\jcp{J.~Chem.~Phys.}%
\def\jgr{J.~Geophys.~Res.}%
\def\jqsrt{J.~Quant.~Spec.~Radiat.~Transf.}%
\def\memsai{Mem.~Soc.~Astron.~Italiana}%
\def\nphysa{Nucl.~Phys.~A}%
\def\physrep{Phys.~Rep.}%
\def\physscr{Phys.~Scr}%
\def\planss{Planet.~Space~Sci.}%
\def\procspie{Proc.~SPIE}%
\let\astap=\aap
\let\apjlett=\apjl
\let\apjsupp=\apjs
\let\applopt=\ao
\bibliographystyle{mn2e}
\bibliography{lit}
\end{document}